\documentclass[12pt]{article}
\usepackage[margin=2.5cm]{geometry}
\usepackage{amsmath}        
\usepackage{amsfonts}       
\usepackage{amssymb}        
\usepackage{graphicx}       
\usepackage{subcaption}     
\usepackage{changepage} 

\usepackage{caption}
\usepackage{xcolor}
\usepackage{booktabs}
\usepackage{hyperref}
\usepackage{footmisc}
\usepackage{appendix}
\usepackage{tikz}
\usetikzlibrary{decorations.pathmorphing, positioning}
\usepackage{float}
\usepackage[numbers,compress]{natbib}

\title{Mutual information-entropy plane: a new quantifier space for  time series analysis}
\author{
    Gaspar Gonzalez Acosta $^{1,2} \footnote{email: \href{mailto:autor@example.com}{gaspar.gonzalez@fisica.unlp.edu.ar}}$, \and Andrés M. Kowalski$^{1,2}$ \\
    \textit{$^{1}$Instituto de Física La Plata, Buenos Aires, Argentina}\\ 
    \textit{$^{2}$Comisión de Investigaciones Científicas (CIC), Buenos Aires, Argentina}
}
\date{\today}

\date{August 2026}

\begin{document}

\maketitle

\begin{abstract}
This work presents a new two-dimensional quantifier plane that combines normalized permutation entropy and normalized permutation mutual information, both derived from Shannon's information theory and normalized consistently through the embedding dimension of the Bandt and Pompe method. Each point on the plane simultaneously characterizes the intrinsic uncertainty of a variable and the amount of information it shares with another. We further show that a third quantity,  equivalent to the conditional entropy of one variable given the other, introduces a notion of informational independence and, when both projections of a variable pair are considered jointly, reveals directionality between them. \\
We analyze regular, chaotic, and stochastic dynamics to illustrate the relevance of the informational plane and show how these regimes evolve as a function of a coupling factor.

\end{abstract}

\section{Introduction}
Entropy and mutual information, both derived from Shannon's information theory, have been extensively used as tools to quantify uncertainty and information in time series arising from dynamical systems of very different nature. In particular, permutation entropy, combined with a statistical complexity measure, has allowed the construction of the so-called complexity-entropy plane, which makes it possible to distinguish between regular, chaotic, stochastic, and white noise dynamics, since chaotic and stochastic systems, despite sharing properties that make them almost indistinguishable through entropy alone, occupy clearly differentiated regions within that plane. This tool has been applied, among other cases, to characterize chaotic maps, self-similar and Gaussian stochastic processes, as well as to classify the dynamics of financial markets and brain activity signals.

Mutual information has been used in fields such as neuroscience \cite{ibanez2020mutual, zbili2021quick}, in financial systems \cite{liu2018generalized,novais2022portfolio}, and in many other areas of knowledge as a tool to quantify the shared information between variables or systems.


In this work, we propose a new quantifier plane that combines the normalized permutation entropy and the normalized permutation mutual information, with the aim of simultaneously characterizing the intrinsic uncertainty of each system and the shared information between them, thus offering a complementary tool for the comparative analysis of complex dynamics. We further show that a third quantity, equivalent to the conditional entropy of one variable given the other, introduces a notion of informational independence and, when both projections of a variable pair are considered jointly, reveals directionality between them. 
We discuss the main advantages of the proposed plane, including its intuitive geometric interpretation, the emergence of a notion of directionality, and its interdisciplinary applicability, as well as its limitations, related to sensitivity to time series preprocessing and the loss of resolution in short or noisy series.

\section{Entropy}
Shannon entropy is defined as follows

\begin{equation}
S\left(P \right)=-\displaystyle \sum_{i=1}^{N} p_{i}
\log\left(p_{{i}}\right),
\label{S}
\end{equation}
where the probability distribution $P=\{ p_{i}\}$ is obtained using different methods, such as histograms and the Bandt-Pompe method. Shannon entropy quantifies the randomness or order of a time series. This means that $S=0$ implies regularity of the system, while $S=\log\left(N\right)$ indicates behavior of maximum disorder \cite{shannon1948mathematical, bandt2002permutation, pessa2021ordpy}.

The normalized entropy takes the following form

\begin{equation}
    H\left(P\right)=\dfrac{S\left(P\right)}{S^{max}}, 
    \label{Hnormal}
\end{equation}
where $S^{max}=\log\left(N\right)$ denotes the entropy corresponding to the uniform distribution. The normalized entropy satisfies the condition $0\leq H\leq 1$.

\section{Mutual Information}
Mutual information ($I$) can be described as follows

\begin{equation}
I(X,Y ) = S(X) + S(Y )-S(X, Y ),
\label{I}
\end{equation}
where $S$ represents the Shannon permutation entropy and $S(X,Y)$ the joint permutation entropy between the variables $X$ and $Y$. In the context of the Bandt and Pompe method, these variables correspond to the ordinal probability distributions, that is,
\[
P^X = \{p_X(\pi_i)\}_{i=1,2,\ldots,d!}, \quad 
P^Y = \{p_Y(\pi_j)\}_{j=1,2,\ldots,d!},
\]
where $\pi_i$ and $\pi_j$ are the ordinal patterns of embedding dimension $d$.

Mutual information is a measure of correlation between $X$ and $Y$, since it quantifies how much information both variables share. In general \cite{rossignoli2013essentials},

\begin{equation}
I(X,Y)\geq 0,
\end{equation}
and it holds that $I(X,Y)=0$ if and only if $X$ and $Y$ are independent, which is equivalent to $S(X,Y) = S(X) + S(Y)$. Conversely, when the variables are totally dependent or indistinguishable ($X=Y$), the joint entropy is $S(X,Y)=S(X)$, and in that case, the mutual information reaches its maximum value,

\[
I(X,X)=S(X),
\]
so that the mutual information (\ref{I}) is bounded below by $0$ (implying non-negativity) and bounded above by \cite{amelio2017correction}
\begin{equation}
\begin{split}
I(X,Y) &\leq \min\{S(X),S(Y)\}\leq \sqrt{S\left(X\right)S\left(Y\right)}\leq \dfrac{1}{2}\big[S(X)+S(Y) \big] \leq\\
 & \textrm{max}\{S(X), S(Y)\} \leq S\left(X,Y\right).
\end{split}
\label{cotas_equations}
\end{equation}

To facilitate comparison between systems, it is convenient to work with normalized mutual information, but which normalization?, because there are several forms depending on the bound chosen (see inequality (\ref{cotas_equations})), which are summarized below \cite{amelio2017correction}
\begin{align}
NMI_1 &=\dfrac{I\left(X,Y\right)}{\textrm{min}\{S\left(X\right),S\left(Y\right)\}},\label{NMI1} \\
MIN_2 &=\dfrac{I\left(X,Y\right)}{\sqrt{S\left(X\right)S\left(Y\right)}}, \label{NMI2} \\
NMI_3&=\dfrac{2I\left(X,Y\right)}{S\left(X\right)+S\left(Y\right)}, \label{NMI3}\\
NMI_4&=\dfrac{I\left(X,Y\right)}{\textrm{max}\{S\left(X\right),S\left(Y\right)\}},\label{NMI4} \\
MIN_5 &=\dfrac{I\left(X,Y\right)}{S\left(X,Y\right)},\label{NMI5}
\end{align}
where $NMI_1$ (\ref{NMI1}) evaluates the degree of saturation of the smaller marginal entropy by the mutual information. This measure reaches unity when all the information of the system with lower entropy is contained in the other, so it is particularly sensitive to functional or near-deterministic dependence relationships (simple systems). $NMI_2$ (\ref{NMI2}) is also known as geometric normalization or as the information correlation coefficient, because it measures dependence based on a geometric mean of the entropies and strongly penalizes relationships where one entropy is very large and the other is very small. $NMI_3$ (\ref{NMI3}) quantifies the average fraction of shared information with respect to the total entropy of both systems. It is symmetric, bounded in $[0,1]$, and reaches its maximum value when $X$ and $Y$ are completely dependent. It is a global symmetric dependence metric. It is the most common form of normalization, especially in the field of clustering and classification, since it treats the total uncertainty of both variables equally as a reference. $NMI_4$ (\ref{NMI4}) measures what proportion of the larger marginal entropy is explained by the mutual information. It is a conservative normalization. A value of $1.0$ is only reached if $I(X, Y) = S(X) = S(Y)$, which is difficult if the marginal entropies differ greatly. The normalization form $NMI_5$ (\ref{NMI5}) measures the proportion of the joint uncertainty that corresponds to information shared between $X$ and $Y$. This normalization directly quantifies the degree of informational redundancy of the joint system, being useful for evaluating how much of the global behavior can be attributed to mutual dependence versus independent variability.

As mentioned previously, in this work we employ the ordinal method of Bandt and Pompe to obtain the probability distribution, therefore, we chose a normalization of the mutual information as a function of the parameter $d$. Since the mutual information reaches its maximum value when $X=Y$, which implies $\textrm{max}\left\{I(X,Y)\right\}=S(X)=S(Y)$, and considering that $X$ and $Y$ have the same dimension, $S(X)\leq S_{max}=\log{(d!)}$ and $S(Y)\leq S_{max}=\log{(d!)}$, for any distribution over $d!$ symbols, this leads to $\max\{S(X),S(Y)\}\leq \log{(d!)}$, therefore, the mutual information with the normalization as a function of the embedding dimension $d$ becomes
\begin{equation}
NMI\left(X,Y \right)=\dfrac{I(X,Y)}{\log{(d!)}}\, \in \left[0,1\right],
\label{MI_alternative}
\end{equation}
which is a particular case of the normalization (\ref{NMI4}) in which one or both entropies reach the absolute theoretical maximum value given by $\log{(d!)}$, which is constant for the system (it only depends on the size of the symbol space); moreover, it is consistent with the normalization of the permutation entropy $H$ (\ref{Hnormal}), since both use the same denominator. When $Y=X$ (maximum relationship between the variables),
\begin{equation}
  \textrm{max}\,   NMI=NMI\left(X,X\right)=\dfrac{S(X)}{\log{(d!)}}=H(X).
  \label{MI_alternative_max}
\end{equation}

By using the normalized permutation entropy $H$ (\ref{Hnormal}) and the normalized permutation mutual information (\ref{MI_alternative}), we are able to construct a two-dimensional information plane in which we can carry out direct comparisons between the variables. This plane arises as a projection of the three-dimensional space of quantifiers, which is originally defined by the triplet \((H(X), H(Y), NMI(X,Y))\).  Each point within the plane allows us to simultaneously characterize (i) the intrinsic uncertainty of the variable $X$ (or $Y$) through entropy, (ii) the amount of information shared between variables X and Y, and (iii) the degree of independence of variable $X$ with respect to variable $Y$ (if the horizontal axis of the plane is $H(X)$).

\section{Mutual information-entropy plane}

From the space defined by the triplet $(H(X), H(Y), NMI(X,Y))$, we propose and analyze a projection onto a two-dimensional mutual information–entropy plane. Through this representation, we provide a relational interpretation that allows for a comparative analysis between dynamical systems.

\subsection{Geometric Interpretation}

This plane contains three quantifiers, namely: normalized mutual information $\text{NMI}(X,Y)$, entropy $H(X)$ (or $H(Y))$, and conditional entropy $H(X|Y)$ (or $H(Y|X)$). $\text{NMI}(X,Y)$ and $H(X)$ are measured on the vertical and horizontal axes, respectively, while $H(X|Y)$ is calculated through the vertical distance between the point $(H(X), NMI(X,Y))$ and the identity line $NMI(X,Y)=H(X)$. Of course, $X$ and $Y$ could be interchanged depending on the desired projection. From the definition of $NMI$, we find that $\Delta_x = H(X) - NMI(X,Y)=H(X|Y)$. That is, the conditional entropy of variable $X$ given variable $Y$. We will use $\Delta_x$ instead of $H(X|Y)$ because of its use as a distance, avoiding confusion given that $H(X|Y)$ is not a distance in probability space. $H(X|Y)$ represents the residual uncertainty of $X$ given $Y$, but, for the purposes of this work, it is convenient to interpret $\Delta_x$ as the ``degree of informational independence" of $X$ with respect to $Y$.

\begin{figure}[H]
 \centering
    \includegraphics[width=0.5\linewidth]{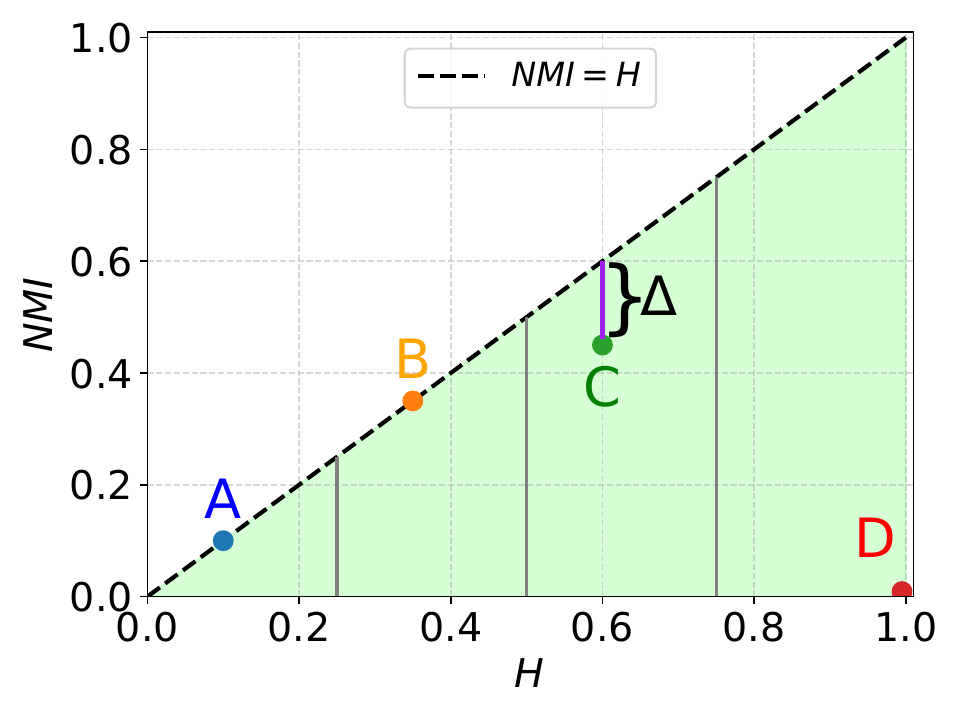}
\caption[Schematic plot of the $NMI$-$H$ plane.]{ Schematic representation of $NMI$ vs. $H$. The green region delimits the admissible zone for the pair of points $(H, NMI)$. The dashed line represents the identity line ($NMI = H$). The vertical distance from a point in the admissible region to that line ($\Delta=H-NMI$) indicates the degree of independence of the variable $X$ relative to $Y$ (if the projection considers the entropy $H(X)$), with maximum independence reached when $\Delta = H$.}
\label{MI_H_referenc}
\end{figure}

When $\Delta_x= 0$, the independence is null, which represents a scenario of maximum informational dependence where knowledge of one variable completely determines the other. Conversely, the upper bound $\Delta_x = H(X)$ corresponds to maximum independence, indicating that the uncertainty about $X$ remains intact given $Y$.

The geometric region colored in green in Figure \ref{MI_H_referenc} is the admissible zone for the pair of points $(H, \text{NMI})$, since for each value of the entropy $H$ the mutual information is bounded by the value $\max \text{NMI}=H$ (see equation (\ref{MI_alternative_max})).
We can divide the $NMI-H$ plane according to 1) the value of $H$: low ($0 \le H < 0.25$), low-medium ($0.25 < H < 0.50$), medium ($H = 0.50$), medium-high ($0.50 < H < 0.75$), and high ($0.75 < H \le 1.00$). This subdivision of entropy ranges serves us to classify values clustered in certain zones of the plane.  2) $\Delta$: considering the projection onto $(H(X), NMI(X,Y))$, so that $\Delta = \Delta_x$, points close to the identity line $NMI=H(X)$ ($\Delta_x \approx 0$) indicate that the variables share almost all their information (strongly synchronized or coupled dynamics); points relatively far from the identity line ($0<\Delta_x<H(X)$) indicate that the variable $X$ is weakly/moderately independent with respect to $Y$, regardless of whether they share little or much information; and points very far from the identity line ($\Delta_x\approx H(X)$) indicate that $X$ has a high degree of independence with respect to $Y$.

But there is a complex relationship among $NMI$, $H$, and $\Delta$. We analyze some points to observe this situation (considering points of the form $P(X,Y)$ and the projection is onto the plane containing the entropy $H(X)$). Points $\text{A}(0.1,0.1)$ and $\text{B}(0.35, 0.35)$ lie on the identity line $\text{NMI}=H(X)$, which indicates null informational independence ($\Delta_x=0$), unlike point $\text{C}(0.6, 0.45)$. However, although point $\text{A}$ lies on the line that indicates maximum informational dependence, the information that the variables share is very small. On the other hand, point $\text{C}$ has greater shared information but a greater degree of independence than point $\text{A}$ (or $\text{B}$), because in the latter the random variable $X$ is completely explained by the random variable $Y$; whereas, at point $\text{C}$ there exists a fraction of the random variable $X$ that cannot be explained by $Y$. At point $\text{C}$, the distance to the identity line $\text{NMI}=H(X)$ is $\Delta_x = 0.6 - 0.45 = 0.15$; therefore, the unexplained fraction of $X$ is $\frac{0.15}{0.6} = 0.25$. That is, $25\%$ of the variable $X$ cannot be explained by $Y$. Point $\textrm{D}(1,0)$ represents the limiting case where one has the maximum theoretical entropy $H(X)=\log(d!)=1$, $NMI=0$, and maximum absolute independence $\Delta_x=H(X)=\log(d!)$.

\subsection{Directionality}
On the same $NMI$–$H$ plane, the point $(H(X), NMI)$ gives, through its vertical distance to the identity line, the conditional entropy $\Delta_x=H(X|Y)=H(X)-NMI$, i.e., how much of $X$ remains unexplained by $Y$. The point $(H(Y), NMI)$, plotted on the same axes, gives instead $\Delta_y=H(Y|X)=H(Y)-NMI$, i.e., how much of $Y$ remains unexplained by $X$. In general, $H(X|Y)\neq H(Y|X)$: the two points tell different stories.

Plotting both points together on the same $NMI$-$H$ plane makes it straightforward to detect which variable is driven by the other (directionality), without resorting to more computationally expensive alternatives such as transfer entropy.

\section{Results: Evaluation and response of quantifiers on synthetic data}

We consider three representative cases of dynamical behavior: regular, 
deterministic chaotic, and stochastic dynamics. The regular dynamics is 
defined as
\begin{equation}
x_t^{\mathrm{reg}} = A\sin(\omega t+\phi),
\end{equation}
where $A$, $\omega$, and $\phi$ denote the amplitude, angular frequency, 
and phase, respectively. Chaotic dynamics is generated using the logistic 
map,
\begin{equation}
x_{t+1}^{\mathrm{chaos}} = r\,x_t^{\mathrm{chaos}}\left(1-x_t^{\mathrm{chaos}}\right), 
\qquad r=4,
\end{equation}
which corresponds to the fully developed chaotic regime. Finally, white 
Gaussian noise is modeled as an independent and identically distributed 
stochastic process,
\begin{equation}
x_t^{\mathrm{WN}} \sim \mathcal{N}(0,\sigma^2).
\end{equation}
Each dynamical realization is normalized to zero mean and unit variance, 
allowing for a direct comparison among the different dynamical regimes.

A second time series is then constructed as
\begin{equation}
y_t = a\,x_t + \sqrt{1-a^2}\,\epsilon_t, \qquad a\in[0,1],
\label{eq:dependent_series}
\end{equation}
where $x_t$ denotes the normalized time series generated by the underlying 
regular, deterministic chaotic, or stochastic dynamics, while $\epsilon_t$ 
is an independent normalized white-noise realization. The parameter $a$ 
controls the relative contribution of the original dynamics to the resulting 
time series. Consequently, assuming statistical independence between $x_t$ 
and $\epsilon_t$, $\operatorname{Var}(y_t)=1$ for all $a\in[0,1]$.

\begin{figure}[H]
 \centering
    \includegraphics[width=0.5\linewidth]{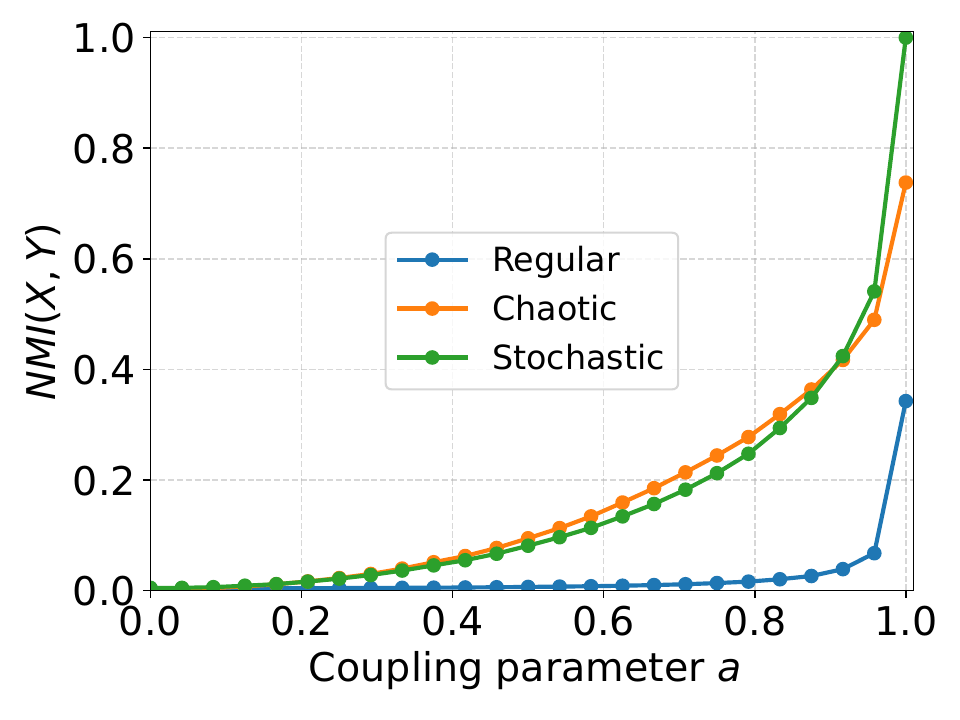}
\caption[$NMI$vs. $a$ .]{Normalized mutual information ($NMI(X,Y)$) vs. the coupling parameter $a$ for the different dynamical regimes.}
\label{NMI_parameter_a}
\end{figure}

Figure~\ref{NMI_parameter_a} shows the normalized mutual information ($NMI(X,Y)$) as a function of the coupling parameter $a$ for the different dynamical regimes under study. As $a \to 1$, the $NMI(X,Y)$ monotonically increases toward its maximum value, indicating complete mutual dependence between the series. Notably, systems with distinct  dynamics can share the same amount of mutual information for a given value of $a$. This behavior highlights the necessity of employing the $NMI(X,Y)$ vs. $H(X)$ plane to properly discriminate among the different dynamical regimes. In the figure, the regular, deterministic chaotic, and stochastic dynamics are depicted by the blue, orange, and green curves, respectively.

\begin{figure}[H]
 \centering
    \includegraphics[width=0.5\linewidth]{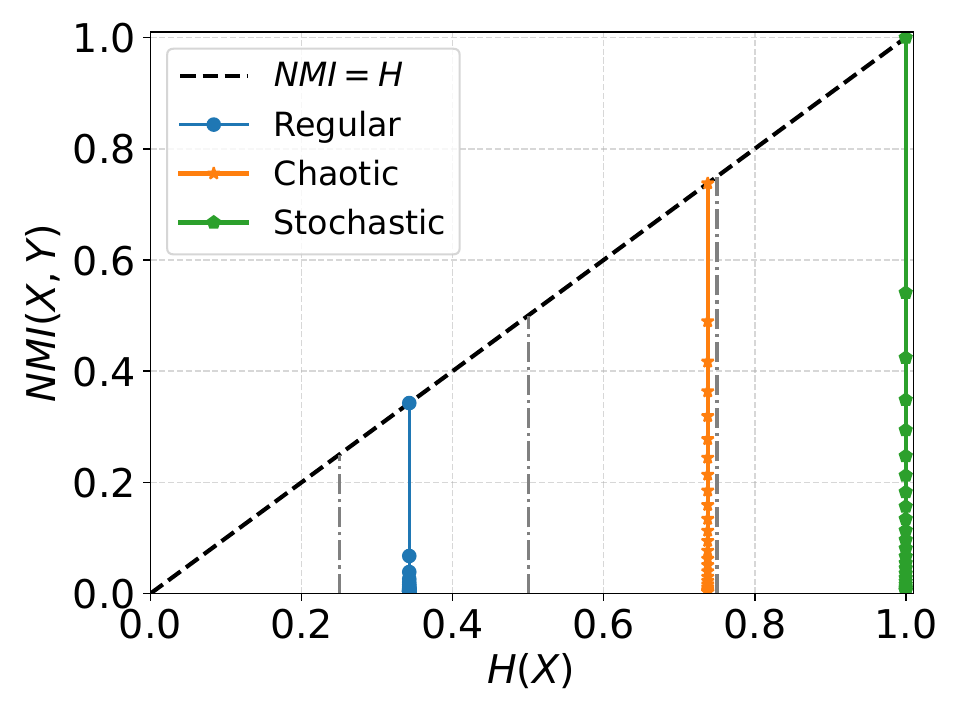}
\caption[$NMI(x_t,y_t)$vs. $H(x_t)$ .]{$NMI(X,Y)$ vs. $H(X)$ for the different dynamics considered.}
\label{NMI_Hx_rcs}
\end{figure}

Figure \ref{NMI_Hx_rcs} shows the graph of $NMI(X,Y)$ versus the normalized permutation entropy $H(X)$ for the different dynamics considered  and  for various values of the coupling parameter $a$.
It can be observed that the entropy remains constant for each type of dynamics, and as the value of $a$ increases, the value of the normalized mutual information $NMI(X,Y)$ increases until it reaches the maximum value given by the identity line $NMI(X,Y)=H(X)$. It is important to note that for this analysis, 25 values of $a \in [0,1]$ were considered, with $NMI = 0$ when $a = 0$ (completely independent dynamics) and $NMI(X,Y) = H(X)$ when $a = 1$ (completely dependent dynamics).

\begin{figure}[H]
 \centering
    \includegraphics[width=0.5\linewidth]{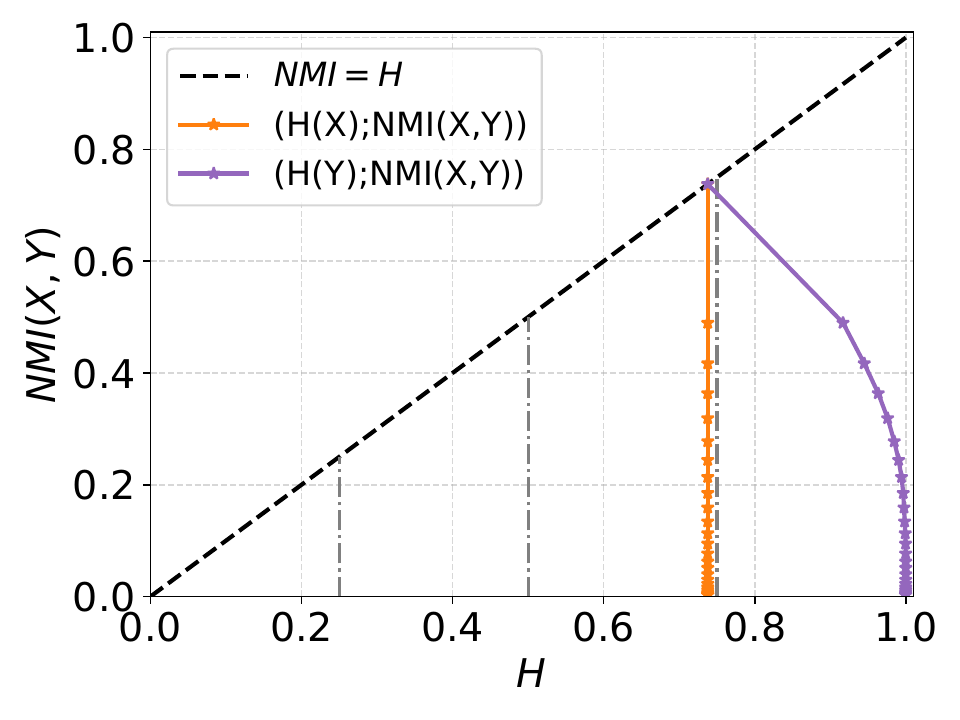}
\caption[$NMI(X,Y)$vs. $H(X)$, $H(Y)$ .]{$NMI(X,Y)$ vs. $H$ for the chaotic dynamics.}
\label{NMI_HxHy}
\end{figure}

Figure~\ref{NMI_HxHy} shows the $NMI(X,Y)$ vs. $H(X)$ and $NMI(X,Y)$ vs. $H(Y)$ planes  to illustrate that projecting onto the entropy of the driving dynamics $x_t$ is generally not equivalent to projecting onto the entropy of the resulting process $y_t$. The points in both representations coincide only under symmetric coupling conditions.
Furthermore, it can be observed that $\Delta_x = H(X) - NMI(X,Y) < \Delta_y = H(Y) - NMI(X,Y)$ across all values of the coupling parameter $a \in [0, 1)$. This asymmetry indicates a clear dependence of $y_t$ on $x_t$, thereby establishing a directional flow of information between the time series.

\section{Discussion}
The mutual information vs. entropy plane is a new tool that we propose for the interpretation of dynamics, and as such, as a tool it has both advantages and disadvantages. Below is a brief descriptive list.
\textbf{Main limitations:}
\begin{itemize}
\item Dependence on time series preprocessing, inherent to statistical quantifiers: The values of entropy and mutual information depend sensitively on the discretization method, the embedding, and the window size used, because both the permutation entropy and the mutual information require a probability distribution, a distribution that is modified according to the amount of available data, window length, embedding dimension, delay, and normalization of the series. Therefore, quantitative comparison between different time series is only valid if all the parameters involved are homogeneous, as we have done in this work.
    
\item Limited resolution for short or noisy series: The mutual information vs. entropy plane can lose discriminating power, since the robust estimation of joint probabilities requires a sufficient number of realizations. This means that for short windows the statistical error can increase, real structural differences are smoothed out, and spurious fluctuations may appear. In these cases, the results should be interpreted as qualitative trends.
\end{itemize}

\textbf{Main advantages:}
\begin{itemize}
\item Simultaneous characterization of internal uncertainty and external dependence: the plane allows analyzing, jointly, the intrinsic uncertainty of a system through entropy $H$ and the shared information through mutual information $NMI$. Their combination offers a richer two-dimensional description than the isolated use of each quantifier.
    
\item Emergence of directionality in the $NMI$-$H$ plane: One of the main benefits of this approach lies in the ability to introduce a directional dimension into the analysis of variables. Although mutual information is inherently symmetric and lacks orientation on its own, its combination with the system's entropy in the $NMI$-$H$ space breaks this symmetry, allowing the geometry of the plane to be interpreted in terms of a direction of influence. 

\item Quantification of the degree of informational independence: The plane provides a direct geometric metric to evaluate the independence between the variables (regardless of how much information they share with each other) through the vertical distance from a point in the admissible region to the identity line, formally defined as $\Delta = H - NMI(X,Y)$. Given that this difference is mathematically equivalent to the conditional entropy $H(X|Y)$ (or $H(Y|X))$, the parameter $\Delta$ measures the remaining uncertainty of one variable given the state of the other; thus, a value of $\Delta =H$ denotes the maximum degree of independence between the variables.

\item Clear and intuitive geometric interpretation: The geometric representation facilitates the qualitative interpretation of the results. As in the complexity-entropy  plan($C_{JS}-H$), regions of the plane can be associated with dynamical regimes (ordered, chaotic, stochastic) and with different degrees of dependence, which is extremely useful for exploratory and comparative analyses.
    
\item Potential for classification and grouping of systems: It provides a natural space for classification, clustering, or dynamic typification tasks. Proximity in the $NMI-H$ plane reflects informational similarities, which makes it possible to identify regional patterns or collective behaviors.
    
\item Transferability and interdisciplinary applicability: the approach is general and transferable to multiple domains. The mutual information vs. entropy plane can be applied to any dynamical system representable through time series, including physical, biological, economic, and social systems, which broadens its potential impact.

\end{itemize}

\section{Conclusions}
In this work, a new quantifier plane, called the $NMI$-$H$ plane, was proposed, combining the normalized permutation entropy and the normalized permutation mutual information into a single two-dimensional representation space. This plane makes it possible to overcome the limitation that arises when these quantifiers are used in isolation: while entropy, on its own, only allows the dynamical regime of an individual system to be characterized, and mutual information, on its own, only quantifies the dependence between two systems without providing information about their intrinsic uncertainty or directionality, the combination of both quantifiers in the $NMI$-$H$ plane enables a simultaneous characterization of both aspects.\\
By evaluating representative cases of regular, deterministic chaotic, and stochastic dynamics, we demonstrated the effectiveness of the $NMI$-$H$ plane in discriminating distinct underlying behaviors and tracking their evolution as a function of coupling.
It was shown that the vertical distance between a point on the plane and the identity line $NMI=H$, defined as $\Delta$, is mathematically equivalent to the conditional entropy $\Delta_x=H(X|Y)$ (or $\Delta_y=H(Y|X)$), which allows the degree of informational independence of one variable with respect to another to be interpreted geometrically: values of $\Delta$ close to zero indicate a strong dependence or synchronization between the variables, while values close to $H$ indicate an almost total independence. Likewise, it was shown that the relationship among $NMI$, $H$, and $\Delta$ is not trivial, since two systems may share the same degree of informational dependence ($\Delta=0$) but differ substantially in the absolute amount of shared information, which evidences the interpretive richness provided by the joint representation of both quantifiers compared to their isolated use. Moreover, since $\Delta_x$ and $\Delta_y$ generally differ, plotting both projections of a variable pair on the same $NMI$-$H$ plane reveals an asymmetry between them, providing a simple geometric notion of directionality, or information flow, without resorting to more computationally expensive alternatives such as transfer entropy.\\
The $NMI$-$H$ plane offers relevant advantages, such as its intuitive geometric interpretation, the emergence of a notion of directionality despite the symmetric nature of mutual information, and its potential application in classification, clustering, and comparison tasks for dynamical systems in fields as diverse as physics, biology, economics, and the social sciences. However, it also presents limitations that must be taken into account when applying it, mainly its sensitivity to time series preprocessing (discretization, embedding, and window size) and the loss of resolution in the case of short or noisy series. Taken together, these results position the $NMI$-$H$ plane as a useful complementary tool for the comparative analysis of uncertainty and dependence in complex dynamical systems.

\bibliographystyle{unsrtnat}

\bibliography{bibliografia}

\end{document}